\documentclass[conference]{IEEEtran}
\IEEEoverridecommandlockouts

\usepackage{tabularx}
\usepackage{booktabs}
\usepackage{caption}
\usepackage{cite}
\usepackage{hyperref}
\usepackage{amsmath,amssymb,amsfonts}
\usepackage{algorithmic}
\usepackage{graphicx}
\usepackage{float}     
\usepackage{textcomp}
\usepackage{xcolor}
\usepackage{caption}
\usepackage{subcaption}
\def\BibTeX{{\rm B\kern-.05em{\sc i\kern-.025em b}\kern-.08em
    T\kern-.1667em\lower.7ex\hbox{E}\kern-.125emX}}
\begin{document}

\title{ XEmoRAG: Cross-Lingual Emotion Transfer with Controllable Intensity Using Retrieval-Augmented Generation  \\
}

\author{
\IEEEauthorblockN{
Tianlun Zuo$^{1}$, Jingbin Hu$^{1}$, Yuke Li$^{1}$, Xinfa Zhu$^{1}$, \\
Hai Li$^{2}$, Ying Yan$^{2}$, Junhui Liu$^{2}$, Danming Xie$^{2}$, Lei Xie$^{1,*}$
}

\IEEEauthorblockA{
$^{1}$Northwestern Polytechnical University, Xi’an, China \\
$^{2}$iQIYI, Inc., Chengdu, China
}
}

\maketitle

\begin{abstract}

Zero-shot emotion transfer in cross-lingual speech synthesis refers to generating speech in a target language, where the emotion is expressed based on reference speech from a different source language.
However, this task remains challenging due to the scarcity of parallel multilingual emotional corpora, the presence of foreign accent artifacts, and the difficulty of separating emotion from language-specific prosodic features.
In this paper, we propose XEmoRAG, a novel framework to enable zero-shot emotion transfer from Chinese to Thai using a large language model (LLM)-based model, without relying on parallel emotional data.
XEmoRAG extracts language-agnostic emotional embeddings from Chinese speech and retrieves emotionally matched Thai utterances from a curated emotional database, enabling controllable emotion transfer without explicit emotion labels. 
Additionally, a flow-matching alignment module minimizes pitch and duration mismatches, ensuring natural prosody. It also blends Chinese timbre into the Thai synthesis, enhancing rhythmic accuracy and emotional expression, while preserving speaker characteristics and emotional consistency.
Experimental results show that XEmoRAG synthesizes expressive and natural Thai speech using only Chinese reference audio, without requiring explicit emotion labels.
These results highlight XEmoRAG's capability to achieve flexible and low-resource emotional transfer across languages.
Our demo is available at \url{https://tlzuo-lesley.github.io/Demo-page/}.

\end{abstract}

\begingroup
\renewcommand\thefootnote{$^{*}$}
\footnotetext{Corresponding author.}
\endgroup


\section{INTRODUCTION}

With the rapid  advancements in the naturalness and expressiveness of text-to-speech (TTS) systems~\cite{ren2020fastspeech,history_chen2021adaspeech,history_liu2021delightfultts,history_kim2021conditional}, research has increasingly focused on multi-emotional and cross-lingual~\cite{cross_huang2022generspeech,cross_zhang2023speak,cross_casanova2024xtts} speech synthesis. This capability is crucial for applications like \textbf{movie dubbing}~\cite{app_biadsy2024zero,app_bigioi2023multilingual,app_liu2023dse}, where synthesized speech must not only convey accurate emotions, but also preserve language-specific prosody and vocal characteristics. 

However, achieving high-quality cross-lingual emotion transfer remains significantly challenging.
First, conventional methods often rely on discrete emotion labels or fixed language identifiers to control affect and linguistic style~\cite{aslp_li2023dicletttsdiffusionmodelbased}. While effective in high-resource settings, these strategies tend to suffer from coarse emotional granularity and poor generalization when applied to \textbf{low-resource languages such as Thai}.
Second, some approaches attempt to disentangle emotional and linguistic representations through learned embeddings or shared intermediate spaces. Although such techniques improve language-independent emotion modeling to some extent, they typically require carefully curated multilingual data pairs and still struggle with transferring prosodic nuances without introducing \textbf{foreign-accented artifacts}. 
Lastly, large language model (LLM)-based models have demonstrated promising results in zero-shot emotion control within individual languages or across linguistically similar language pairs. However, their ability to generalize across typologically distant and phonetically diverse languages, such as Chinese to Thai, remains limited 
These gaps underscore the need for more effective frameworks that can achieve controllable, natural, and emotionally expressive speech synthesis across languages in zero-shot scenarios.

To address these challenges, we propose \textbf{XEmoRAG}, a novel framework that leverages emotion retrieval and flow-matching alignment to generate expressive Thai speech from Chinese emotional references. Our method does not require parallel data or explicit emotion labels, enabling natural and controllable cross-lingual emotion transfer. By effectively addressing resource scarcity and prosodic mismatches, XEmoRAG is well-suited for real-world applications demanding high-quality emotional speech synthesis.

To enable effective zero-shot emotion transfer, we propose a \textbf{retrieval-augmented generation (RAG)} mechanism that automatically selects emotionally similar prompts from an external database of high-quality speech samples spanning diverse emotions. Retrieval-based prompting enables the model to adapt emotional expressions across languages while naturally incorporating prosodic features of Thai, such as pitch, duration, and intonation—thereby avoiding the common issue of foreign-accented speech that occurs when prosody is directly transferred between languages. Furthermore, we introduce emotion intensity control to allow flexible modulation of emotional strength during synthesis.
To address the scarcity of emotion-labeled data in low-resource languages like Thai, we adopt a two-stage fine-tuning approach using Llasa~\cite{ye2025Llasa}, a state-of-the-art large language model designed for multilingual speech synthesis and understanding tasks. In the first stage, the model is fine-tuned on abundant non-emotional Thai speech data to learn the foundational phonetic and linguistic characteristics unique to Thai, effectively building its base capability. In the second stage, with this solid foundation, the model requires only a small amount of emotional Thai speech data to learn to express emotions accurately. Combined with our novel RAG mechanism, this approach effectively overcomes data scarcity, enabling precise cross-lingual emotion transfer and outperforming traditional models like DelightfulTTS~\cite{liu2021delightfultts} in generating natural and intelligible speech for low-resource languages.
To further improve prosody and naturalness, we incorporate \textbf{flow-matching (FM)} alignment to reduce pitch and duration mismatches between the source and target languages. FM promotes expressive and natural prosody while adapting to the unique characteristics of the target language. Moreover, it integrates the timbre from the original Chinese prompt, preserving the distinct vocal qualities of the target language and preventing a robotic tone.
Experimental results show that our approach outperforms baseline models in generating more natural and expressive cross-lingual emotional speech, especially in Chinese-to-Thai emotional movie dubbing, demonstrating its strong practical value.

\section{RELATED WORK}

\subsection{Cross-Lingual Emotion Transfer Speech Synthesis}
Recent years have witnessed remarkable advances in emotional speech synthesis \cite{emo_zhang2023iemottsrobustcrossspeakeremotion}\cite{aslp_li2023zero}.
In the field of emotional speech synthesis, emotion-ID-based methods are widely adopted~\cite{emo_zhang2023iemottsrobustcrossspeakeremotion}\cite{emoid_kang2023zetspeechzeroshotadaptiveemotioncontrollable}. These approaches generate emotionally expressive speech by conditioning the model on discrete emotion labels (e.g., happy, sad). While effective when ample annotated data are available, such methods often produce over-smoothed emotional expressions due to the discrete nature of the emotion labels. They also struggle to generalize to low-resource languages, failing to capture subtle emotional variations across languages. 

In cross-lingual scenarios, language-ID-based conditioning is employed to control language-specific prosody and rhythm\cite{language_zhang2019learning}\cite{language_zhang2023speakforeignlanguagesvoice}. By providing the model with a language identifier during inference, these approaches aim to separate linguistic information from speaker identity, facilitating the generation of natural cross-lingual synthetic speech. However, this method alone is insufficient to eliminate foreign accent artifacts, as different languages possess distinct phonemes and phonetic features. Consequently, relying solely on language IDs may not fully address the challenge of achieving native-like pronunciation and emotional expressiveness in the synthesized speech. 

To address these limitations, recent studies have explored more sophisticated techniques. For instance, DiCLET-TTS \cite{aslp_li2023dicletttsdiffusionmodelbased} leverages diffusion probabilistic models to facilitate cross-lingual emotion transfer. It disentangles speaker and emotion features via orthogonal projections, enhancing expressiveness and reducing accent problems.
METTS~\cite{aslp_zhu2023mettsmultilingualemotionaltexttospeech} encodes bilingual text into a unified symbol space and uses multi-scale emotion modeling to separate coarse-grained (language-invariant) and fine-grained (language-specific) prosody, but it depends on paired multilingual data and is not designed for retrieval-based or zero-shot transfer.
The emergence of large-scale speech foundation models has significantly expanded the frontier of controllable speech synthesis, particularly in emotional expressiveness~\cite{llm_du2025cosyvoice3inthewildspeech,llm_wang2024maskgctzeroshottexttospeechmasked,llm_zeng2024glm4voiceintelligenthumanlikeendtoend,llm_zhang2023speakforeignlanguagesvoice}.
For example, CosyVoice 3~\cite{llm_du2025cosyvoice3inthewildspeech} enables fine-grained intra-lingual emotion transfer via prompts but lacks explicit cross-lingual emotion modeling.
Similarly, MaskGCT~\cite{llm_wang2024maskgctzeroshottexttospeechmasked} supports multilingual emotion control through discrete tokens and performs well in zero-shot settings. Nevertheless, it lacks explicit mechanisms for disentangling and transferring emotional expressions across typologically distant languages, and its reliance on discrete control codes may constrain the granularity and naturalness of emotional migration. 
These limitations highlight the gap between multilingual emotion control and genuine cross-lingual emotion transfer. 
Leveraging the strong generalization capacity and prompt-following ability of large decoder-only language models, we are motivated to explore whether such models can bridge this gap. 

\subsection{Retrieval-Augmented Generation for Cross-Lingual Emotion Transfer}

Retrieval-Augmented Generation (RAG) was originally proposed in Natural Language Processing (NLP) ~\cite{rag_lewis2021retrievalaugmentedgenerationknowledgeintensivenlp} to enhance generation by retrieving semantically relevant external knowledge and conditioning outputs on it. By integrating parametric generation with non-parametric memory, RAG improves factual accuracy and contextual coherence in knowledge-intensive tasks.

Recently, RAG-inspired frameworks have been explored in speech-related tasks such as style-guided synthesis~\cite{rag_luo2025autostyle} and audio-conditioned dialogue generation~\cite{rag_chen2025wavragaudiointegratedretrievalaugmented}. For instance, AutoStyle-TTS\cite{rag_luo2025autostyle} introduces a retrieval-based automatic style matching mechanism that selects prosodically and semantically appropriate speech samples from a style library to guide synthesis. It leverages multiple embedding extractors to capture style features and retrieve reference utterances, resulting in more natural and expressive synthesized speech. Similarly, WavRAG\cite{rag_chen2025wavragaudiointegratedretrievalaugmented} proposes a retrieval-augmented generation framework that directly operates on raw audio inputs, bypassing the need for intermediate automatic speech recognition (ASR) transcriptions and thereby mitigating the risk of recognition errors. By incorporating a WavRetriever component, WavRAG enables cross-modal retrieval from a mixed text-audio knowledge base, significantly improving contextual reasoning in speech dialogue generation.

However, current retrieval-augmented TTS approaches predominantly focus on monolingual scenarios and lack effective mechanisms for zero-shot cross-lingual emotion transfer, where discrepancies arise from language differences and mismatches in emotions. This challenge is particularly significant in low-resource settings, such as Chinese-to-Thai emotion transfer, where parallel bilingual emotional speech data from the same speaker is unavailable. Under these conditions, simply retrieving stylistically similar utterances from the source language often results in prosodic misalignment or foreign-accented speech in the target language. Moreover, existing frameworks generally do not provide controllability over emotional intensity, which is critical for applications like multilingual movie dubbing.
 
\section{METHODS}
\label{sec:method}

\begin{figure*}[t]
    \centering
    \includegraphics[width=0.9\linewidth]{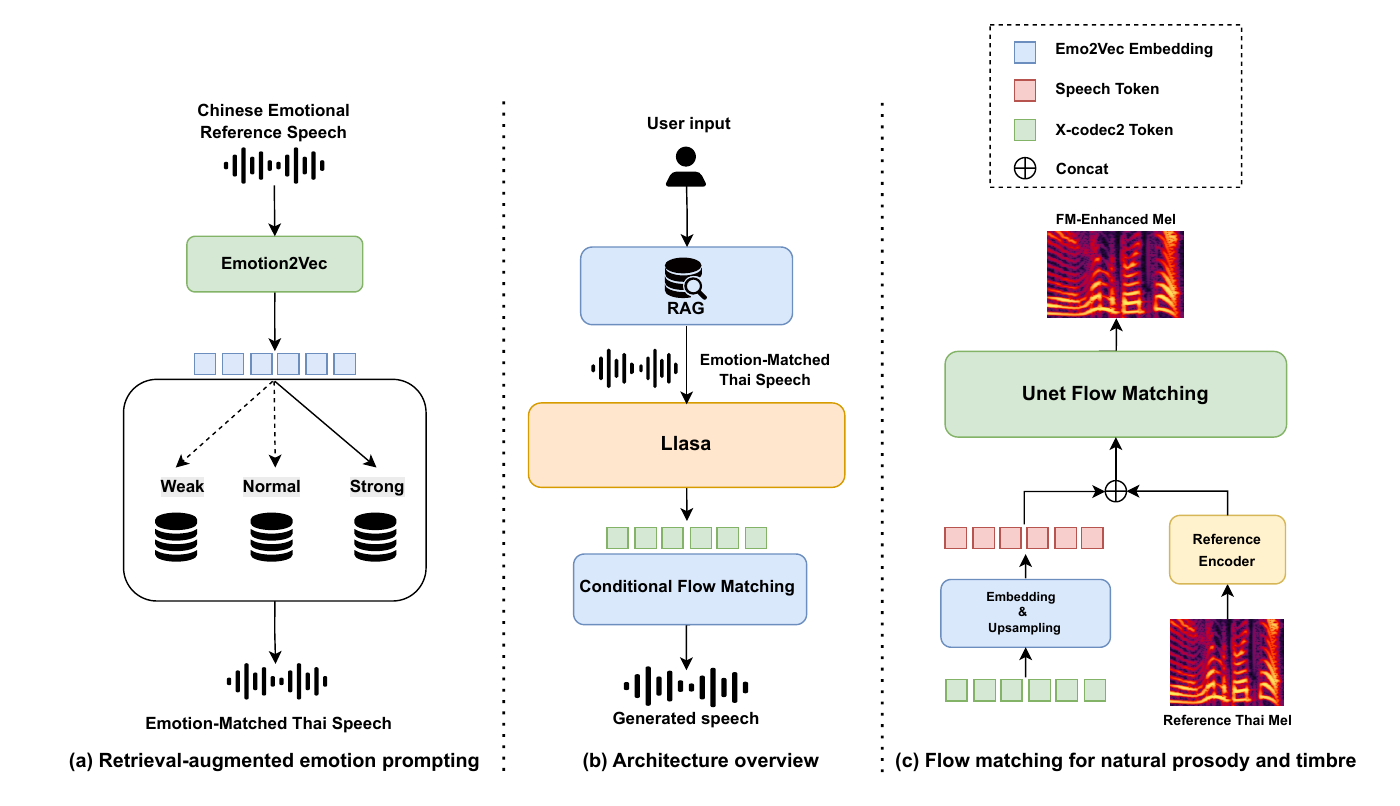}
    \caption{Cross-lingual emotional speech synthesis system based on retrieval-augmented generation and flow matching. The user input includes a Chinese emotional reference speech, target Thai text, and a specified emotion intensity level (strong, normal, or weak).}
    \label{fig:overview}
\end{figure*}

An overview of our proposed cross-lingual emotional speech synthesis framework is illustrated in Fig.~\ref{fig:overview}. 
The system is designed to generate emotionally expressive Thai speech by leveraging a Chinese reference utterance, integrating powerful tokenization, alignment, and retrieval-based style control modules to achieve natural and controllable synthesis across languages without parallel emotional data.

\subsection{Overview}
Our framework synthesizes emotionally expressive Thai speech from Thai text by transferring the emotional style conveyed in a Chinese reference speech, without relying on any parallel emotional data.
The backbone of zero-shot cross-lingual emotion transfer speech synthesis follows the Llasa~\cite{ye2025Llasa}, a Transformer-based text-to-speech framework. 
By operating in an autoregressive next-token prediction manner, it replaces traditional multi-stage TTS pipelines with a unified, decoder-only model that predicts discrete acoustic tokens, enabling end-to-end speech generation within the LLM framework. We use the Llasa-1B-Multilingual version, a 1B-parameter model trained on multilingual speech data, which allows for flexible and natural synthesis across multiple languages.

Speech waveforms are first encoded into discrete token representations using X-Codec2~\cite{ye2025codec}, a powerful tokenizer that fuses both semantic and acoustic features of speech via a single-layer vector quantizer. 
These discrete tokens, paired with corresponding text inputs, form the joint input-output space that the Transformer learns to model.
Lastly, to further improve the prosodic naturalness of synthesized speech, we incorporate flow-matching alignment, which improves the mapping between input text and acoustic tokens, thereby reducing pitch and duration mismatches during generation.
For emotion transfer, we introduce a RAG module that selects emotionally aligned Thai utterances based on emotional representations extracted by the Emo2Vec pre-trained model\cite{wang2020Emo2Vec}.
These retrieved utterances serve as prosodic prompts to guide the emotional style during generation. Additionally, the retrieval process is conditioned on emotion intensity levels—weak, normal, or strong—thus enabling precise control over the expressive strength of the synthesized output.

Together, these components form a comprehensive architecture that enables the cross-lingual generation of natural, emotionally rich, and controllable Thai speech from Chinese input.

\subsection{Flow-matching Alignment and Speech Reconstruction}

Flow matching utilizes continuous normalizing flows (CNFs) to align the distributions of token sequences by learning vector fields that trace the target mel-spectrogram.
Within the framework of token-to-mel speech synthesis, this approach utilizes an ordinary differential equation (ODE) solver to generate mel-spectrogram samples that closely align with the desired data distribution.

To achieve this, we adopt a transformer-based 1D U-Net architecture as the vector field predictor within the flow matching framework. This model integrates time embeddings, residual down/up-sampling blocks, and 16-head self-attention mechanisms to enhance the global modeling of speech signals.We designed the structure based on this to address the following two key issues.

Firstly, to address the frame rate mismatch between speech tokens and mel features, we upsample the speech tokens by a factor of \(1.6:1\) using linear interpolation, aligning their frame rate with that of the mel-spectrograms.

Meanwhile, to further enhance generation diversity and speaker control, we condition the model on speaker reference embeddings, aiming to strengthen the modeling capability of timbral characteristics across source and target speakers.

Thus, we use flow matching based on the convolutional Transformer UNet to map discrete token representations generated by the X-Codec2 tokenizer, enriched with semantic and acoustic information, to mel-spectrograms representing the target speech representations.And the UNet parameters are optimized by minimizing the L1 loss between the predicted vector and the corresponding ground truth.

This flow-based alignment strategy reduces pitch and duration mismatches during synthesis, thereby improving prosodic naturalness. By jointly modeling semantic content, acoustic structure, and speaker timbre, our approach supports high-quality and controllable speech reconstruction, making it well-suited for emotional speech synthesis in multilingual scenarios.

\subsection{Retrieval-Augmented Generation for Emotion Transfer}

We integrate the RAG module into our framework to facilitate zero-shot cross-lingual emotion transfer.
Given a reference speech in Chinese that conveys emotion implicitly without explicit emotion labels, we first extract a language-agnostic emotional embedding using a pre-trained Emo2Vec model. Next, we perform similarity-based retrieval over a curated pool of Thai emotional utterances to identify the sample that best matches the reference emotion. To enhance retrieval diversity and enrich the emotional expressiveness of the synthesized speech, we employ a clustering-based retrieval strategy. Specifically, we apply K-means clustering to the emotional embeddings extracted by Emo2Vec from the Thai utterance pool, using a distance metric that reflects emotional proximity. The retrieved utterance then serves as a prosody prompt during inference, enabling the model to generate Thai speech that faithfully preserves the emotional characteristics of the original Chinese reference without producing unnatural foreign accents.

\begin{figure}[t]
  \centering
  \begin{subfigure}[b]{0.48\textwidth}
    \centering
    \hspace{0.03\linewidth}
    \includegraphics[width=0.86\linewidth]{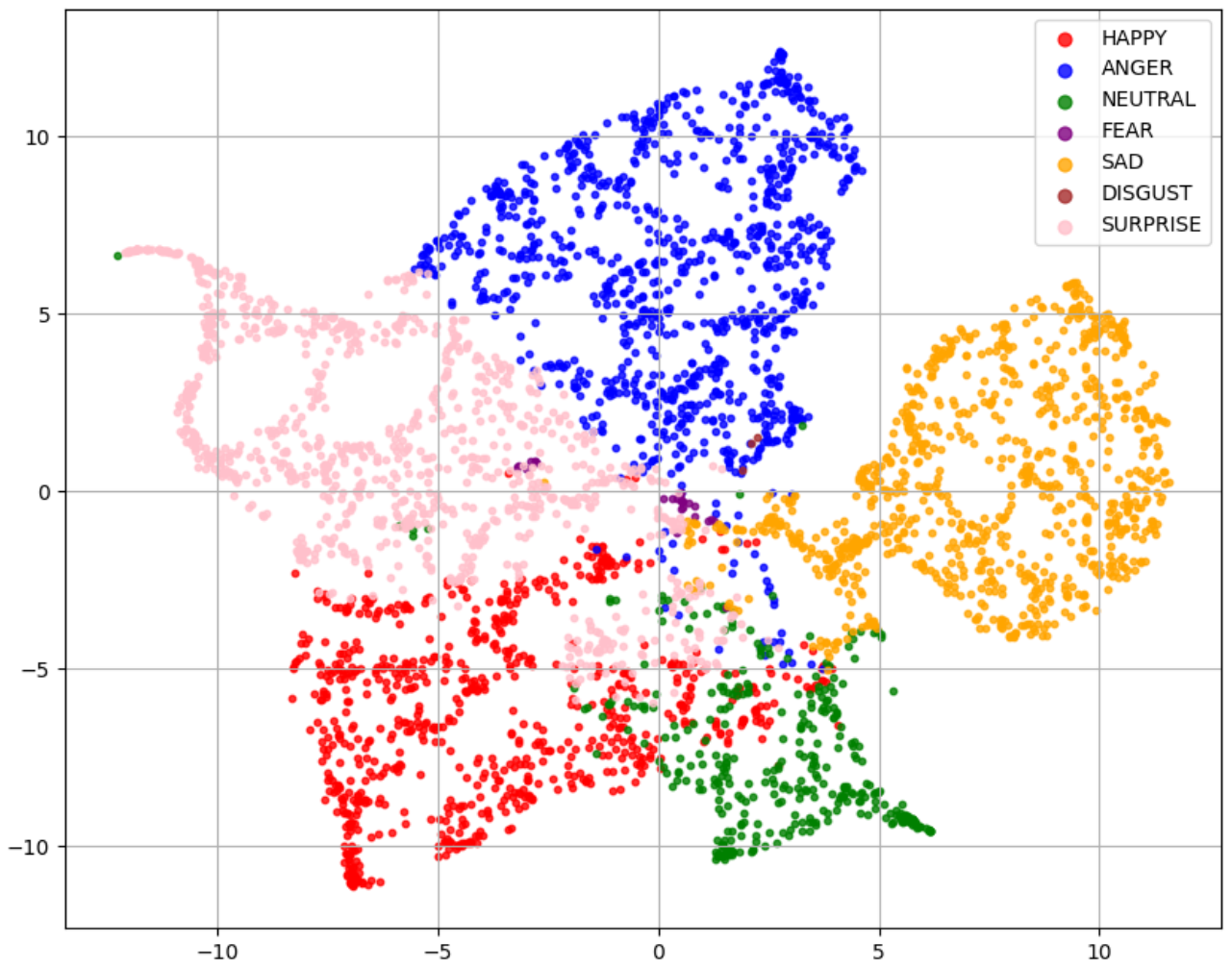}
    \caption{Mixed Chinese-Thai Emotional Embeddings (Emo2Vec)}
    \label{fig:mix}
  \end{subfigure}
  \hfill
  \begin{subfigure}[b]{0.47\textwidth}
    \centering
    \hspace{0.03\linewidth}
    \includegraphics[width=0.88\linewidth]{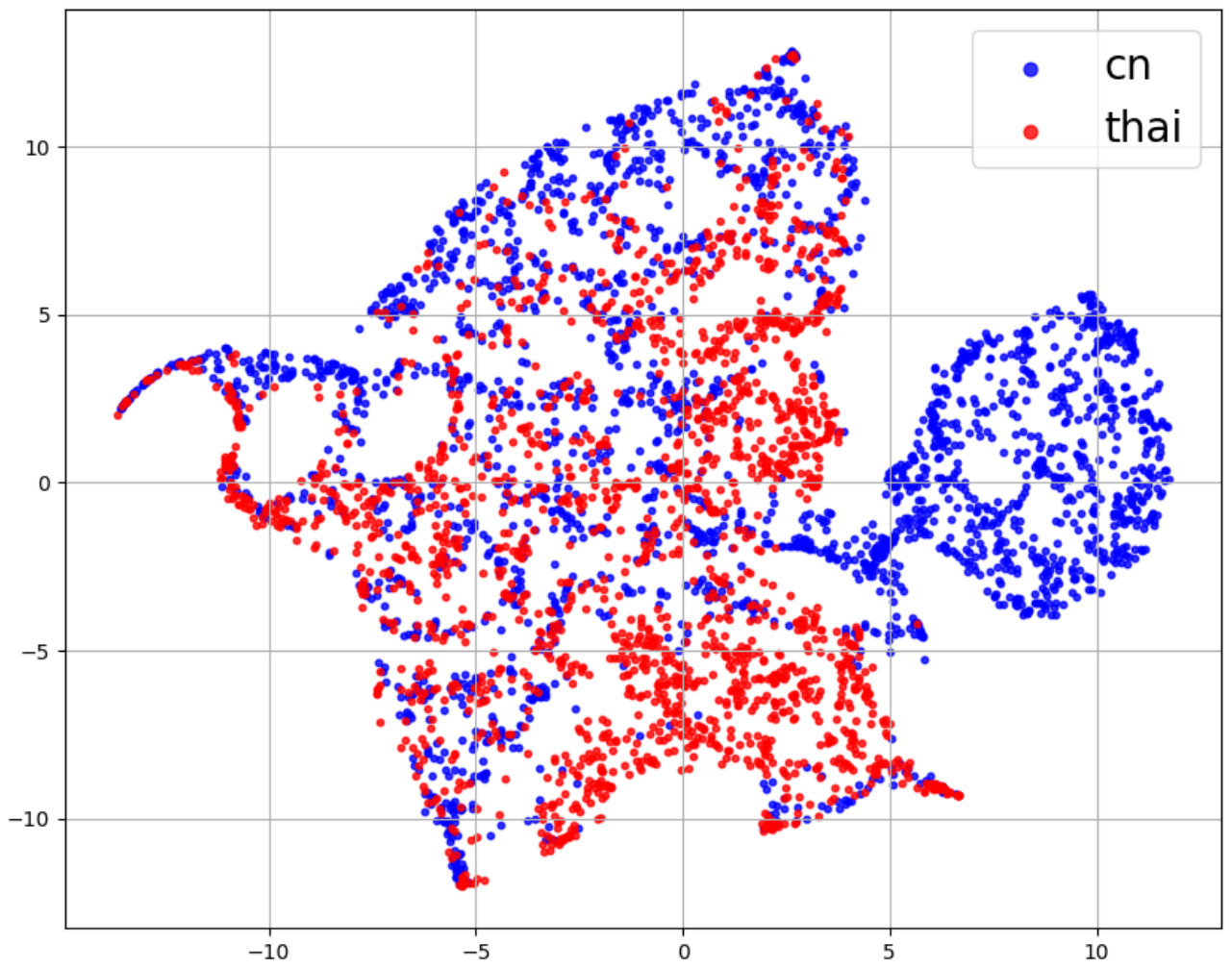}
    \caption{Language-Specific Distributions (Chinese vs Thai)}
    \label{fig:split}
  \end{subfigure}
  \caption{t-SNE visualization of emotional representations extracted by Emo2Vec.}
  \label{fig:combined}
\end{figure}

To demonstrate the cross-lingual generalizability of the emotional embeddings derived from the Emo2Vec model, we visualize their distributions using t-SNE plots. As illustrated in Fig.~\ref{fig:mix}, when Chinese and Thai utterances are jointly projected, their emotional embeddings form well-aligned clusters, suggesting a language-agnostic structure. Furthermore, Fig.\ref{fig:split} shows the emotion clusters separately for Chinese and Thai utterances, with similar shapes and boundaries, indicating the robustness of Emo2Vec embeddings across languages. This cross-lingual consistency justifies the use of these embeddings for emotion-based retrieval in a zero-shot setting.

To further enhance controllability, we introduce emotion intensity levels into the retrieval process. In our Thai dataset, emotional utterances are annotated with three levels of intensity: weak, normal, and strong. 
A desired intensity level can be specified at inference time, which guides the RAG module to restrict retrieval to only those utterances in the corresponding emotional subset. For instance, when a strong emotional intensity is specified, the retrieval is limited to the \textit{strong} subset only.

This strategy enables not only accurate cross-lingual emotion transfer but also provides multi-level control over the emotional strength of the generated speech. It helps synthesize Thai speech that better matches the intended expressive style of the reference, while preserving prosodic naturalness and avoiding unnatural foreign accents.

\subsection{Training and Inference}

During training, we adopt a unified modeling approach where Thai discrete speech tokens, extracted by a neural codec named X-Codec2, are concatenated with their corresponding text tokens to form the training sequences. The model is trained to predict each speech token conditioned on both the input text tokens and the previously generated speech tokens, without relying on explicit speaker or emotion labels. This design ensures that the model can generalize to unseen speakers, languages, and emotional styles in a zero-shot fashion.

Significantly, the RAG module is only activated during inference. As shown in Fig.~\ref{fig:overview}, given a reference utterance in the source language, we first extract its emotional embedding using a pre-trained Emo2Vec model. Based on this embedding, it retrieves the most emotionally similar Thai emotion embedding from a curated database. 
The retrieved Thai speech audio and its corresponding text transcription are provided as prosodic and semantic prompts, respectively. These, along with the target Thai text, serve as inputs to the Llasa model for speech generation. The Llasa model generates discrete speech tokens in an autoregressive manner. These tokens are subsequently passed through a flow-matching model to obtain mel-spectrograms, which are then converted into waveforms using DSPGAN~\cite{song2023dspgan}.

This inference pipeline enables zero-shot, cross-lingual, controllable emotion transfer, producing speech that is both emotionally expressive and prosodically natural in the target language.

\section{EXPERIMENTS}

\subsection{Datasets}

The backbone model, Llasa-1B-Multilingual~\cite{ye2025Llasa}, is pre-trained on a large-scale multilingual speech corpus containing approximately 80K hours of speech from over 40 languages, including low-resource and tonal languages. This extensive pre-training enables strong cross-lingual generalization and robust speech synthesis capabilities, making it a powerful foundation for downstream tasks such as cross-lingual emotional speech synthesis.

We first fine-tune the Llasa model with 1K hours of internal non-emotional Thai speech data to strengthen its ability to capture the phonetic characteristics of the Thai language. 
Following this, a second fine-tuning stage is conducted on the Thai-optimized model using 60 hours of internal emotional speech data. 
All speech data is down-sampled to 16 kHz, and 80-dimensional mel-spectrograms are extracted with a 12.5 ms frame shift and a 50 ms frame length, ensuring high-quality acoustic features for both training and inference tasks.

\subsection{Model Configuration}

We adopt the Typhoon2-Audio 8B Instruct model~\cite{pipatanakul2024typhoon} and DelightfulTTS~\cite{liu2021delightfultts} as the compared systems for zero-shot cross-lingual emotion transfer speech synthesis.

\begin{itemize}
    \item 
    \textbf{Typhoon2-Audio 8B Instruct}: 
    A unified, end-to-end large-scale model with billions of parameters, covering the entire process from speech understanding to generation. It supports traditional TTS as well as complex tasks such as cross-lingual emotion transfer. Optimized for Thai, it is well-suited for Thai text-to-speech synthesis.
    \item 
    \textbf{DelightfulTTS}: 
    The DelightfulTTS~\cite{aslp_li2023zero} implementation uses a Non-Autoregressive Predictive Coding module for accent modeling 
    and hierarchical HuBERT embeddings for cross-lingual emotion transfer.
    
\end{itemize}

For the Llasa training setup, the network is optimized using the Adam optimizer with an initial learning rate of \(1e-5\) and a batch size of 16. The model is trained for 25K steps.

For the flow matching training setup, 
we use the AdamW optimizer with an initial learning rate of \(1e-5\), a 1000-step warmup, and StepLR decay scheduling, with a batch size of 32 over 1 million training steps.

\begin{table*}[t]
\caption{Comparison of the proposed model with baseline systems and ablation models. Evaluation metrics include emotion similarity MOS (EMOS), naturalness MOS (NMOS), character error rate (CER), and speaker cosine similarity (SIM).}
\renewcommand{\arraystretch}{1.1}
\begin{center}
\begin{tabular}{lcccc}
\toprule
\textbf{Model} & \textbf{EMOS $\uparrow$} & \textbf{NMOS $\uparrow$} & \textbf{CER (\%) $\downarrow$} & \textbf{SIM $\uparrow$} \\
\midrule
DelightfulTTS & 3.89 ± 0.12 & \textbf{4.41 ± 0.14} & 7.33 & 0.83 \\
Typhoon2-Audio 8B Instruct & - & - & 8.72 & - \\
\cmidrule(r){1-5}
XEmoRAG (proposed) & \textbf{4.65 ± 0.10} & 4.38 ± 0.12 & \textbf{5.95} & \textbf{0.89} \\
 w/o flow-matching & 4.20 ± 0.11 & 3.76 ± 0.13 & 6.17 & 0.74 \\
 w/o RAG Module & 4.25 ± 0.10 & 3.99 ± 0.12 & 6.73 & - \\
\bottomrule
\end{tabular}
\label{table:comparison}
\end{center}
\end{table*}

\subsection{Evaluation of XEmoRAG and Baseline Models}
We evaluate the proposed XEmoRAG model alongside two baselines, DelightfulTTS and Typhoon2-Audio 8B Instruct, using both subjective and objective metrics.

Subjective evaluation employs Mean Opinion Score (MOS) tests on emotion similarity and naturalness, rated by fifteen native listeners (10 Chinese and 5 Thai).
The evaluation data consist of emotional speech from a single Chinese speaker covering diverse emotions, along with a Thai emotional speech dataset annotated with multiple emotion categories (e.g., happiness, sadness, fatigue and anger). Notably, the emotion labels in the Thai dataset are unseen during inference to realistically simulate cross-lingual emotional transfer. The system takes Chinese speech as input and synthesizes Thai speech conveying the same emotional content without using explicit emotion annotations.
As shown in Table~\ref{table:comparison}, the proposed XEmoRAG model achieves the highest Emotion Similarity MOS of 4.65, significantly outperforming DelightfulTTS (3.89), demonstrating superior cross-lingual emotional expressiveness. In terms of Naturalness MOS, XEmoRAG scores 4.38, closely matching DelightfulTTS's 4.41, indicating that emotional enhancement does not compromise speech naturalness.

For objective evaluation, we compute speaker cosine similarity (SIM) using embeddings extracted by a pre-trained WeSpeaker~\cite{wang2023wespeaker} model and character error rate (CER) using a Zipformer-based ASR system~\cite{yao2023zipformer}. As shown in Table~\ref{table:comparison}, XEmoRAG achieves the highest SIM score of 0.89, which surpasses all baselines and reflects strong preservation of speaker identity.
Additionally, it achieves the lowest CER of 5.95\%, indicating improved intelligibility and synthesis quality compared to DelightfulTTS (CER 7.33\%) and Typhoon2-Audio 8B Instruct (CER 8.72\%).

\subsection{Evaluation of Retrieval Methods and Database Sizes}

To assess the impact of retrieval strategies and database sizes on accuracy and latency, we conduct controlled experiments comparing embedding-based and clustering-based retrieval across two database scales: 3K and 8K utterances.
In contrast to the clustering-based retrieval strategy, the embedding-based method directly computes cosine similarity between the reference emotion embedding and all candidates, selecting the one with the highest similarity as the emotion-matched Thai prompt.
For each setup, we measure retrieval accuracy which is defined as the proportion of retrieved utterances matching the Chinese reference emotion and average inference time per utterance, covering retrieval and synthesis. Accuracy is evaluated on a manually labeled test set, while inference time is averaged over all samples.

As shown in Table~\ref{tab:retrieval_comparison}, cosine similarity yields slightly higher accuracy at 3K utterances (84.1\%) than clustering (82.7\%), indicating better reliability in small, well-distributed search spaces. However, at 8K utterances, cosine similarity accuracy drops to 82.4\% and inference latency rises sharply to 2.85s per utterance. Conversely, clustering scales efficiently, achieving the highest accuracy of 86.3\% at 8K with only 1.67s latency. This highlights structured retrieval’s advantage in handling large emotional pools without sacrificing speed or quality. These results justify our use of clustering-based retrieval as the default in RAG, balancing accuracy, speed, and scalability in cross-lingual synthesis.

Based on these findings, we adopt clustering-based retrieval with the 8K database as our default, balancing accuracy and speed to ensure high-fidelity, efficient cross-lingual emotional synthesis.
\begin{table}[b]
\centering
\caption{Retrieval accuracy and inference time under different retrieval strategies and database sizes.}
\label{tab:retrieval_comparison}
\renewcommand{\arraystretch}{1.2} 
\resizebox{\columnwidth}{!}{%
\begin{tabular}{lcccc}
\toprule
\textbf{Retrieval Method} & \textbf{Database Size} & \textbf{Accuracy (\%)} & \textbf{Inference Time (s)} \\
\midrule
Embedding-based Retrieval        & 3K                     & 84.1           & 1.32                         \\
Embedding-based Retrieval        & 8K                     & 82.4           & 2.85                         \\
Clustering-based Retrieval       & 3K                     & 82.7           & \textbf{1.18}                \\
Clustering-based Retrieval       & 8K                     & \textbf{86.3}  & 1.67                         \\
\bottomrule
\end{tabular}%
}
\end{table}

\subsection{Ablation study}

To assess the individual contributions of key components within our proposed framework, we conduct a series of ablation experiments focusing on the flow-matching alignment mechanism and the RAG module.
First, we investigate the impact of replacing Llasa's original X-Codec2-based alignment with our flow-matching strategy. The original Llasa system utilizes X-Codec2 for alignment during decoding. In contrast, our approach employs a flow-matching alignment designed to provide more accurate and flexible temporal alignment between the source and target speech features. This substitution enables a direct comparison between the two alignment techniques. Experimental results reveal that adopting flow-matching leads to significant improvements in both audio quality and emotional expressiveness, highlighting its superiority over the original alignment method.
As detailed in Table~\ref{table:comparison}, removing the flow-matching mechanism causes the EMOS score to decrease from 4.65 to 4.20, while SIM drops sharply from 0.89 to 0.74. This indicates that flow-matching plays a vital role in maintaining emotional nuance and preserving speaker identity in the synthesized speech.

Additionally, we examine the effect of the RAG module by evaluating the model performance without this component. The absence of the RAG module results in a reduction of the EMOS score to 4.25, reflecting a noticeable decline in the emotional expressiveness and intelligibility of the generated speech. This demonstrates that retrieval-augmented prompting effectively enhances the model’s ability to generate nuanced and natural emotional speech.

Together, these ablation studies confirm that both the flow-matching alignment and the RAG module are essential components. Their integration significantly boosts the perceptual quality and objective metrics of synthesized speech, validating the design choices of our framework for zero-shot cross-lingual emotional speech synthesis.

\section{CONCLUSION}

In this paper, we propose XEmoRAG, a cross-lingual emotional speech synthesis framework that generates natural and expressive Thai speech from Chinese inputs with controllable emotion intensity. 
Our method eliminates the need for parallel emotional data or explicit emotion labels by leveraging a large-scale multilingual backbone, flow-matching alignment, and retrieval-based emotion prompting.
Experimental results demonstrate its effectiveness and practical potential in scenarios such as movie dubbing.
Given the language-agnostic nature of the extracted emotion representations, our framework can be readily adapted to other languages. Future work will explore extending beyond the current Chinese-to-Thai setting to support a wider range of cross-lingual emotion transfer scenarios.

\bibliographystyle{IEEEbib}
\bibliography{arxiv}

\vspace{12pt}

\end{document}